# Visualization of unidirectional optical waveguide using topological photonic crystals made of dielectric material

Yuting Yang[1], Yun Fei Xu[1], Tao Xu[1], Hai-Xiao Wang[1], Jian-Hua Jiang[1], Xiao Hu[2*] and Zhi Hong Hang[1*]

[1]*College of Physics, Optoelectronics and Energy & Collaborative Innovation Center of Suzhou Nano Science and Technology, Soochow University, Suzhou 215006, China*
[2]*International Center for Materials Nanoarchitectonics (WPI-MANA), National Institute for Materials Science, Tsukuba 305-0044, Japan*
*e-mail: HU.Xiao@nims.go.jp; zhhang@suda.edu.cn

**The introduction of topology unravels a new chapter of physics.[1-3] Topological systems provide unique edge/interfacial quantum states which are expected to contribute to the development of novel spintronics and open the door to robust quantum computation.[4-9] Optical systems can also benefit from topology. Engineering locally in real space a honeycomb photonic crystal with double Dirac cone in its photonic dispersion, topology transition in photonic band structure is induced and a pseudospin unidirectional optical channel is created and demonstrated by the backscattering immune electromagnetic transportation. The topological photonic crystal made of dielectric material can pave the road towards steering light propagations and contribute to novel communication technology.**

By paralleling Maxwell's equations for harmonic modes of electromagnetic waves and Schrödinger equation for electrons, Haldane and Raghu[10] revealed that chiral edge states can be realized for electromagnetic waves, which was confirmed experimentally using gyro-magnetic materials under external magnetic field[11-13] and helical optical fibers with light injection[14]. Topology protected electromagnetic propagations with time-reversal symmetry (TRS) are preferred as being more compatible with established electronics technology. However, when TRS is preserved, the difference between electromagnetic waves and electronic systems becomes more crucial. Photons, the quantum version of electromagnetic wave, obey Bose-Einstein statistics, while electrons follow Fermi-Dirac statistics and carry intrinsically the Kramers pair associated with the spin-down and -up states, which is crucial for the helical edge states in quantum spin Hall effect (QSHE).[4-8] Recent effort has been devoted to generate TRS topological electromagnetic propagations[15], with a variety of ways to mimic Kramers doubling in bosonic systems. Bianisotropic materials[16-17], coupled resonator wave guides[18-19], structures with engineered synthetic gauge field[20-22] and waveguides with broken mode symmetry[23-24] have been proposed and experimentally realized.

Photonic crystals (PCs) are analog of electronic crystals with atoms replaced by periodic arrangement of permittivity and permeability and the interaction is through light rather than Coulomb force.[25] Different from limited type of atoms available for building electronic crystals, "artificial atoms" for PCs have large configurability not to mention possible complexity of the lattice. In this Letter, topologically protected unidirectional optical channel is created using two-dimensional (2D) PC structures made of conventional dielectric materials, as shown in Fig. 1a. A hexagonal cluster of six neighboring dielectric rods forms the "artificial atom" and "artificial atoms" are arranged in a triangular lattice. There are two competing lengths in such a composite lattice, as schematically shown in Fig. 1c: the intra couple distance $h_1$ defined as the distance between neighboring rods within the same cluster and the inter couple distance $h_2$ defined as the distance between the nearest rods in neighboring clusters and $2h_1+h_2=a$, where $a$ is the lattice constant of the triangular lattice of "artificial atoms". When $h_1=h_2$, a Dirac cone dispersion with two-fold degeneracy exists at the Brillouin zone center (Fig. 1b), which is equivalent to a honeycomb lattice. Doubly degenerate Dirac cone dispersion with four-fold degenerate states is the key to mimic QSHE with pseudospin and spin-orbital coupling in photonic system.[26] Deforming the honeycomb cluster slightly in a designed way, *i.e.* the ratio between $h_1$ and $h_2$, opens a gap in the Dirac photonic bands at Γ point and PCs with inversed bulk band properties can be created. The band inversion can be understood from electromagnetic conductivity in photonic conduction bands. As schematically shown in Fig. 1c, in $PC_1$ with $h_1>h_2$, the inter cluster coupling is strong and electromagnetic wave can easily transport across clusters. When $h_1<h_2$ as in $PC_2$, electromagnetic waves have much more weights confined within individual clusters. The detailed discussion about photonic band mode properties and band inversion can be found in Supplementary Information. By assembling $PC_1/PC_2$ with inverted bulk band properties in the upper/lower side of the interface, topological interfacial states are expected.[27] Because the present recipe is formulated based on conventional dielectric materials, such as silicon and alumina, this mechanism for realizing topological photonic state can be applied to visible lights, where gyro-magnetic materials are unavailable and loss in metals becomes serious. A similar topological transition was observed experimentally in an acoustic system, with the photon replaced by phonon, which is induced by tuning the radius of steel cylinders.[28]

The photonic band diagram of a super-cell of $PC_1$ and $PC_2$ calculated by COMSOL Multiphysics is shown in Fig. 2a. Each "artificial atom" is composed of six identical alumina cylinders, with diameter $d=6mm$ and relative permittivity $\varepsilon=7.5$. The polarization under consideration is with the electric field along the cylinder axis. The lattice constant is $a=25mm$. The intra (inter) couple distances are $h_1=0.36a/0.30a$ ($h_2=0.28a/0.40a$) in $PC_1/PC_2$ respectively. The ratio between $h_1$ and $h_2$ is different in the two PCs, yielding different bulk band properties, which guarantee the existence of two topological interface states in the common gap between $PC_1$ and $PC_2$, as highlighted in green and purple respectively in Fig. 2a. From eigenmode analysis of the interface states, the local Poynting vector distributions of the purple and green interface states are rotating anti-clockwise and clockwise in the *xy* plane, respectively, which, defined as $\Phi_\uparrow$ and $\Phi_\downarrow$, represent the two different pseudospin photonic states (see supplementary information for details). One important feature of the present approach is to generate pseudospin and thus Kramers doubling for electromagnetic modes based on the $C_{6V}$ crystal symmetry for the purpose of achieving topological states of light with TRS. In order to demonstrate this feature in a direct way, we design a four-antenna array where the phase of electromagnetic wave is tuned to increase/decrease by 90 degree clockwise between neighboring antennas by using delay lines, which generates $\Phi_\uparrow/\Phi_\downarrow$. This makes the demonstration of helical edge states characterizing QSHE to a level unavailable in any other systems, including both electronic, photonic and phononic systems explored so far. As highlighted in the inset of Fig. 1a, the antenna array is put on the interface of $PC_1/PC_2$ at the position $x=0$ in Fig. 2b. The experimental visualization of $\Phi_\uparrow$ propagation is shown in Fig. 2b at 7.42GHz, where the electromagnetic field is much enhanced in the forward direction, consistent with its positive group velocity seen in Fig. 2a. Moreover, within the common photonic bandgap of $PC_1$ and $PC_2$, the electromagnetic wave can only transport along the interface between $PC_1$ and $PC_2$ (at $y=0$ in Fig. 2b). The profile of electric field magnitude at $x=100$mm is plotted in Fig. 2d, which decays into $PC_1$ and $PC_2$, with decay length 47.2mm and 43.7mm respectively.

In the present experiment, time-harmonic field distributions at all frequencies of interest can be measured and thus the transmission to forward/backward directions can be experimentally evaluated by doing line integral on profile of electric field intensity. The resultant transmissions to S1 and S2 as indicated by dashed lines in Fig. 2b are displayed in Fig. 2c. Three different frequency regions can be identified. Region I corresponds to the frequency range within the common bandgap between $PC_1$ and $PC_2$. The transmission in the forward direction (S1) is larger than that in the backward direction (S2) by over one order of magnitude. In this frequency region, the electromagnetic field is well confined in the interfacial region, as typically shown in Figs. 2b and 2d. In Region II where the frequency is below the bulk band edge around 7.35GHz, the interface state gradually gets closer to the bulk photonic modes of $PC_1/PC_2$ as the frequency decreases, where influences from the bulk modes are enhanced. In electronic systems, characteristic unidirectional transportation of topological edge/interface state is diminished as long as its energy level overlaps with bulk states.[5-6] However, in photonic systems, the pseudospin unidirectional propagation can survive the interference from bulk bands. Remarkably, as shown in Fig. 2c, the forward directional propagation still overwhelms the backward one in Region II. The transmission to forward and backward directions increase simultaneously and their difference gradually decreases upon frequency, which provides a side proof that our antenna with given pseudospin does not emit electromagnetic waves in any preferred direction, and thus the unidirectional propagation shown in Fig. 2b is purely of topology origin. The effect can also be visualized from the field distributions in Region II, with the profile at $x=100$mm and 7.04GHz given in Fig. 2e where the local field in bulks is much stronger as compared to in Region I (see Fig. 2d), indicating that certain amount of bulk modes in $PC_1$ and $PC_2$ are excited. Nevertheless, the interface state still survives since the field magnitude is maximal at the interface and decays into both PCs. In Region III of frequency, starting from 7.55GHz where the frequency is above the upper band edge, the interface state is influenced strongly by the bulk photonic bands. We notice that in Region III the group velocities of the interface states are very small and even reversed away from the Γ point. The measured transmissions in forward and backward directions quickly become comparable.

In order to verify the effectiveness of pseudospin design and the backscattering-immune property of the topological optical channel created, another experiment is carried out as displayed in Fig. 3a. The region with $PC_1$ is highlighted in orange, whereas the remaining region is of $PC_2$ structure. The antenna array now is designed to excite $\Phi_\downarrow$ pseudospin state. The measured field distribution at the frequency of 7.408GHz is presented in Fig. 3b. Even though the total propagation length is nearly three times of that as given in Fig. 2b, and a 120 degree sharp turn exists on the interface, the field magnitude of the interface state remains unchanged at the exit port. The transmission obtained from field intensity integration (S2) is nearly identical to that at positions close the source (S1) as shown in Fig. 3c, indicating that the topological interface state with given pseudospin transports without noticeable back-scattering from the sharp turn.

In conclusion, unidirectional backscattering immune optical pathway is demonstrated with proof-of-principle microwave experiments. An antenna array is designed as the source of electromagnetic wave with given pseudospin polarization, and the propagation of pseudospin polarized state in electromagnetic systems is visualized. Even in the frequency bands where the topological edge state is buried in bulk photonic band edges, directional transmission is still achieved. Because only dielectric materials with reasonable permittivity are adopted, the structure can be readily available at near-infrared or even optical wavelengths[29-30] and the present design of topological photonic crystal will contribute to future optical communication technology.

# Reference

[1] Klitzing, K. V., Dorda, G. & Pepper, M. New method for high-accuracy determination of the fine structure constant based on quantized Hall resistance. *Phys. Rev. Lett.* **45**, 494-497 (1980).

[2] Thouless, D. J., Kohmoto, M., Nightingale, M. P. & den Nijs, M. Quantized Hall conductance in a two-dimensional periodic potential. *Phys. Rev. Lett*. **49**, 405–408 (1982).

[3] Haldane, F. D. M. Model for a Quantum Hall Effect without Landau Levels: Condensed-Matter Realization of the "Parity Anomaly". *Phys. Rev. Lett*. **61**, 2015 (1988).

[4] Kane, C. L. & Mele, E. J. Quantum spin Hall effect in graphene. *Phys. Rev. Lett.* **95**, 226801 (2005).

[5] Bernevig, B. A., Hughes, T. L. & Zhang, S. C. Quantum spin Hall effect and topological phase transition in HgTe quantum wells. *Science* **314**, 1757-1761 (2006).

[6] König, M. et al. Quantum Spin Hall Insulator State in HgTe Quantum Wells. *Science* **318**, 766-770 (2007).

[7] Hasan, M. Z. & Kane, C. L. Colloquium: topological insulators. *Rev. Mod. Phys.* **82**, 3045-3067 (2010).

[8] Qi, X.-L. & Zhang, S.-C. Topological insulators and superconductors. *Rev. Mod. Phys.* **83**, 1057-1110 (2011).

[9] Weng, H.-M., Yu, R., Hu, X., Dai, X., & Fang, Z. Quantum Anomalous Hall Effect and Related Topological Electronic States. *Adv. Phys.* **64**, 227- 282 (2015).

[10] Haldane, F. D. M. & Raghu, S. Possible realization of directional optical waveguides in photonic crystals with broken time-reversal symmetry. *Phys. Rev. Lett.* **100**, 013904 (2008).

[11] Wang, Z., Chong, Y., Joannopoulos, J. D. & Soljačić, M. Observation of unidirectional backscattering immune topological electromagnetic states. *Nature* **461**, 772-775 (2009).

[12] Poo, Y., Wu, R., Lin, Z., Yang, Y. & Chan, C. T. Experimental realization of self-guiding unidirectional electromagnetic edge states. *Phys. Rev. Lett*. **106**, 093903 (2011).

[13] Skirlo, S., Lu, L., Igarashi, Y., Yan, Q., Joannopoulos, J. D., Soljačić, M. Experimental Observation of Large Chern numbers in Photonic Crystals, *Phys. Rev. Lett.*, **115**, 253901 (2015)

[14] Rechtsman, M. C. et al. Photonic Floquet topological insulators. *Nature* **496**, 196-200 (2013).

[15] Lu, L., Joannopoulos, J. D., and Soljačić, M. Topological states in photonic systems, *Nature Physics*, **12**, 626-629 (2016)

[16] Khanikaev, A. B. et al. *Nature Materials* **12**, 233–239 (2013)

[17] He, C. et al. Photonic topological insulator with broken time-reversal symmetry. *Proc. Natl Acad. Sci. USA* **113**, 4924–4928 (2016).

[18] Hafezi, M., Demler, E. A., Lukin, M. D. & Taylor, J. M. Robust optical delay lines with topological protection. *Nature Physics*. **7**, 907-912 (2011).

[19] Hafezi, M., Mittal, S., Fan, J., Migdall, A. & Taylor, J. M. Imaging topological edge states in silicon photonics. *Nature Photon.* **7**, 1001-1005 (2013).

[20] Fang, K., Yu, Z. & Fan, S. Realizing effective magnetic field for photons by controlling the phase of dynamic modulation. *Nature Photon.* **6**, 782-787 (2012).

[21] Rechtsman, M. C. et al. Strain-induced pseudomagnetic field and photonic Landau levels in dielectric structures. *Nature Photon.* **7**, 153–158 (2013).

[22] Tzuang, L. D., Fang, K., Nussenzveig, P., Fan, S. & Lipson, M. Non-reciprocal phase shift induced by an effective magnetic flux for light. *Nature Photon.* **8**, 701-705 (2014).

[23] Chen,W.-J. et al. Experimental realization of photonic topological insulator in a uniaxial metacrystal waveguide. *Nature Commun*. **5**, 6782 (2014).

[24] Cheng, X. et al. Robust reconfigurable electromagnetic pathways within a photonic topological insulator *Nature Materials* **15**, 542–548 (2016).

[25] Joannopoulos, J. D., Johnson, S. G., Winn, J. N., & Meade, R. D., Photonic Crystals: Molding the Flow of Light, 2nd ed. (Princeton University Press, Princeton, NJ, 2008).

[26] Wu, L.-H. & Hu, X. Scheme for achieving a topological photonic crystal by using dielectric material. *Phys. Rev. Lett*. **114**, 223901 (2015).

[27] Xiao, M, Zhang, Z. Q. & Chan C. T. Surface Impedance and Bulk Band Geometric Phases in One-Dimensional Systems, *Phys. Rev. X* **4**, 021017 (2014).

[28] He, C. et al. Acoustic topological insulator and robust one-way sound transport, *Nature Physics* doi:10.1038/nphys3867 (2016).

[29] Gabrielli, L., Cardenas, J., Poitras, C. B., Lipson, M. Silicon nanostructure cloak operating at optical frequencies, *Nature Photon*. **3**, 461-463 (2009).

[30] Moitra, P. et al. Realization of an all-dielectric zero-index optical metamaterial, *Nature Photon*. **7**, 791-795 (2013).

**Figure Captions:**

**Figure 1 | Design principle of topological photonic crystal. a,** Experimental setup photo to realize pseudospin photonic states and unidirectional optical channel. Triangle arrays of 4x12 "artificial atoms" made of six dielectric cylinders with parameters of $PC_1$ (in pink) and $PC_2$ (in blue) are assembled in the regions above and below the interface (green line). A square shaped antenna array (red circled and enlarged in inset) is used to selectively excite one of the two pseudospin photonic states. The whole setup is surrounded by absorbing materials to eliminate unwanted scattering, as shown in blue. **b**, Doubly degenerate Dirac dispersion can be found at the Brillouin zone center when the intra couple distance $h_1$ and inter couple distance $h_2$ are equal to each other, which yields the honeycomb lattice referring to individual dielectric cylinders, as guaranteed by the $C_{6v}$ lattice symmetry. **c**, Schematics of the distribution of Poynting vector in one of the two pseudospin states when the intra couple distance $h_1$ and inter couple distance $h_2$ are competing. The electromagnetic conductivity will be largely different due to the ratio between $h_1$ and $h_2$: $h_1/h_2>1$ in $PC_1$ whereas $h_1/h_2<1$ in $PC_2$.

**Figure 2 | Experimental verification of pseudospin unidirectional propagation. a**, Calculated band diagram of a super-cell of $PC_1$ and $PC_2$. The actual super-cell used in simulations is highlighted in grey and periodic (absorbing) boundary condition is applied on $x$ ($y$) respectively. The lattice constant is common for the two PCs $a=25$mm and the two different "artificial atoms" are made of the same alumina dielectric rods with diameter $d=0.24a=6$mm and the relative permittivity $\varepsilon=7.5$. The intra and inter unit cell coupling distances are $h_1=0.36a$, $h_2=0.28a$ in $PC_1$ and $h_1=0.30a$, $h_2=0.40a$ in $PC_2$. Dispersions of $\Phi_\uparrow$ and $\Phi_\downarrow$ pseudospin interface states are shown in purple and green respectively. The folded bulk bands of $PC_1$ and $PC_2$ are shadowed in grey. **b**, Experimentally measured $E_z$ field distribution at 7.42GHz of the sample setup as in Fig. 1a, where a unidirectional propagation is clearly observed. The antenna array is designed to excite $\Phi_\uparrow$ state and the photonic frequency and momentum are indicated by the triangle in Fig. 2a. **c**, Measured transmission to the two ends S1/S2 of the sample, indicated by the dash lines in Fig. 2b. Three different frequency regions can be observed. Region I corresponds to the frequency range within the common gap between $PC_1$ and $PC_2$. In Region II where the frequency is below the lower bulk band edge, even though the interface states interact with the bulk PC modes, unidirectional propagation can still be observed. In Region III where the frequency is above the upper bulk band edge, the difference between forward and backward transmission is quickly diminished due to the non-monotonic group velocity of interface states. **d-e**, Profiles of field magnitude distributions along the $y$ direction at $x=100$mm away from the source with frequency 7.42GHz and 7.04GHz indicated by the triangle and square in Fig. 2c respectively.

**Figure 3 | Experimental verification of backscattering immune optical channel with sharp turn. a**, Experimental setup photo for the system with a 120 degree sharp turn in the interface between $PC_1$ and $PC_2$, where a pseudospin state $\Phi_\downarrow$ is generated from the antenna, as shown in red dashed circle. The region with $PC_1$ is highlighted in orange whereas the remaining region is of $PC_2$ structure. Because $PC_1$ is now below the interface (opposite to that in Fig. 1a), it is $\Phi_\downarrow$ state which propagates forwardly. All the parameters are the same as those given in Fig. 2. **b**, Measured

$E_z$ field distribution at frequency of 7.408GHz. **c**, Measured transmission to S1/S2 in the frequency range inside Region I.

# Figures

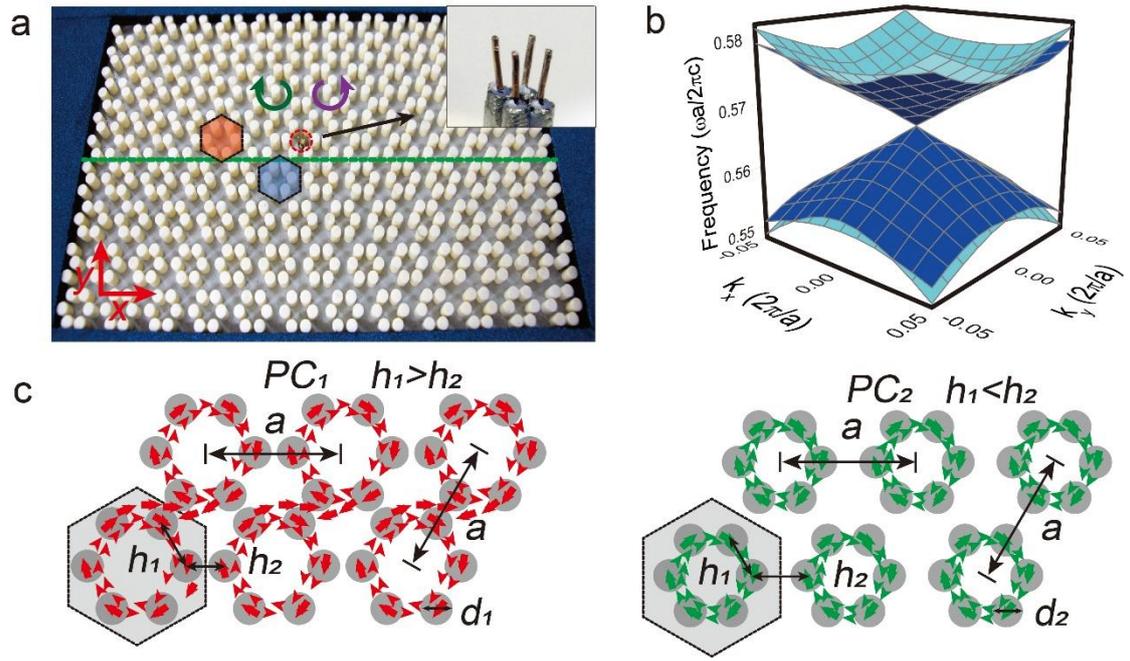

**Figure 1**

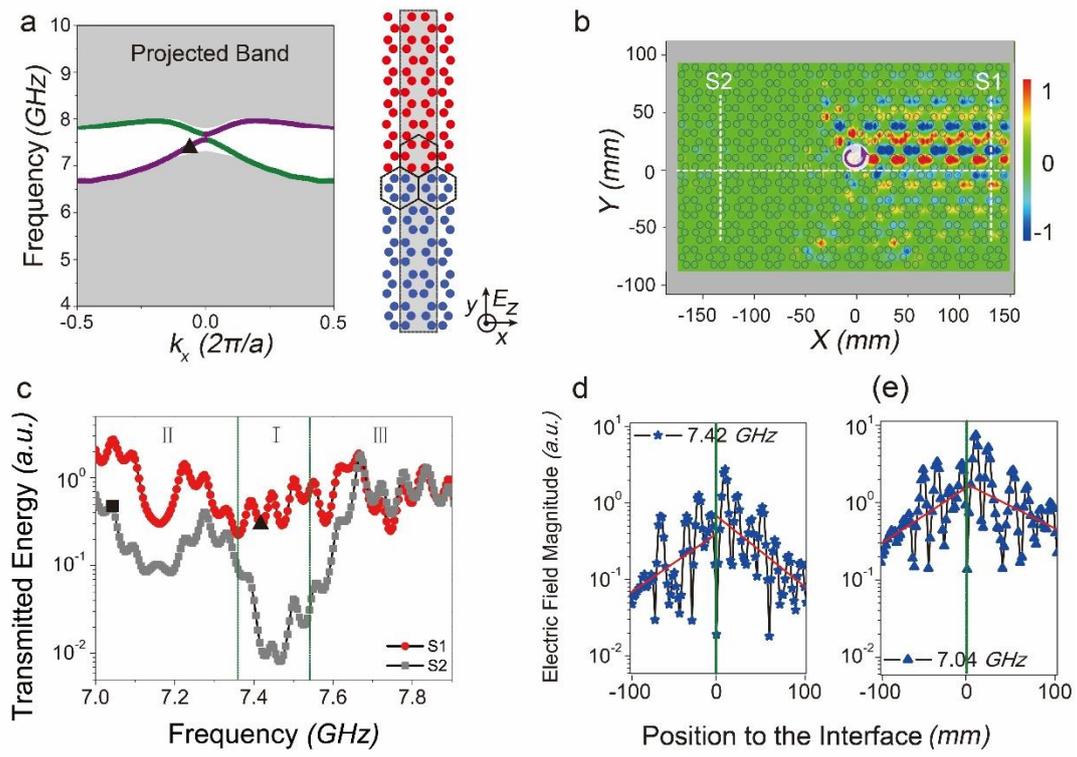

**Figure 2**

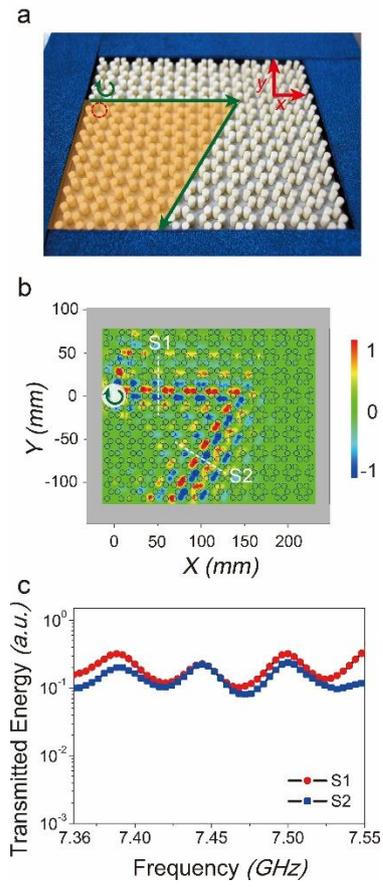

**Figure 3**